\documentclass[twocolumn,superscriptaddress,prl]{revtex4}
\UseRawInputEncoding
\usepackage{amsmath,amsfonts,amssymb}
\usepackage{pdfpages}
\usepackage{graphicx}
\usepackage{xcolor}
\usepackage{xr-hyper}
\usepackage[colorlinks=true,urlcolor=blue,citecolor=blue,linkcolor=blue]{hyperref}
\begin{document}

\title{Drude weight and the many-body quantum metric in one-dimensional Bose systems}
\author{G. Salerno}
\email{grazia.salerno@aalto.fi}
\affiliation{Department of Applied Physics, Aalto University School of Science, FI-00076 Aalto, Finland}
\author{T. Ozawa}
\affiliation{Advanced Institute for Materials Research (WPI-AIMR), Tohoku University, Sendai 980-8577, Japan}
\author{P. T\"{o}rm\"{a}}
\email{paivi.torma@aalto.fi}
\affiliation{Department of Applied Physics, Aalto University School of Science, FI-00076 Aalto, Finland}
\affiliation{Advanced Institute for Materials Research (WPI-AIMR), Tohoku University, Sendai 980-8577, Japan}
\begin{abstract}
We study the effect of quantum geometry on the many-body ground state of one-dimensional interacting bosonic systems. We find that the Drude weight is given by the sum of the kinetic energy and a term proportional to the many-body quantum metric of the ground state. Notably, the many-body quantum metric determines the upper bound of the Drude weight. We validate our results on the Creutz ladder, a flat band model, using exact diagonalization at half and unit densities. Our work sheds light on the importance of the many-body quantum geometry in one-dimensional interacting bosonic systems.
\end{abstract}
\maketitle

The geometrical properties of a physical system's eigenstates, defined over a parameter space, are encoded in the quantum geometric tensor~\cite{resta2011insulating, kolodrubetz2017geometry, rossi2021quantum}. The real and the imaginary part of such a tensor are respectively the quantum metric (also known as Fubini-Study metric) and the Berry curvature. 
Recent studies have found that the quantum metric of Bloch bands plays a role in many phenomena, including the non-adiabatic anomalous Hall effect and the orbital magnetic susceptibility~\cite{gao2014field, piechon2016geometric, iskin2018spin, gianfrate2020measurement}, entanglement effects~\cite{kuzmak2018entanglement}, topology in 2D Chern insulators \cite{ozawa2021relations, mera2021kahler, mera2021engineering}, or the topological charge of inflated Dirac monopoles~\cite{salerno2020floquet}. 

In the case of flat-band systems~\cite{kopnin2011high, leykam2018artificial}, the role of the quantum metric is essential to describe the superfluid properties of interacting fermions~\cite{peotta2015superfluidity, tovmasyan2016effective, liang2017band, huhtinen2022revisiting}. 
While no superfluidity in flat bands would naively be expected due to an infinite effective mass of single particles, even a weak interaction is able to create mobile pairs~\cite{torma2018quantum}. The mobility of pairs and a finite superfluid weight are given by the minimal quantum metric of the Bloch states defined in momentum space~\cite{peotta2015superfluidity,torma2018quantum,huhtinen2022revisiting}. This mechanism has inspired plenty of further work in the field of flat band superconductivity and beyond, and may be relevant for twisted-bilayer graphene where nearly-flat bands arise~\cite{hu2019geometric, julku2020superfluid, xie2020topology, peri2021fragile,torma2022superconductivity,kruchkov2022quantum,tian2023evidence, kruchkov2023quantum}. Also in bosonic systems, the quantum metric defined in momentum space is known to be related to the speed of sound and the excitation fraction of Bose-Einstein condensates on flat bands within the Bogoliubov theory~\cite{julku2021excitations, julku2021quantum, subacsi2022quantum, iskin2023quantum}. Nonetheless, recent work has highlighted the limitations of the Bogoliubov theory in assessing the superfluid weight~\cite{julku2022superfluidity}, and Popov's hydrodynamic theory approaches have been applied~\cite{kruchkov2023unconventional}.

All these previous works relating the superfluid and other physical properties to the quantum metric have been limited to the contributions of quantum metric defined with single-particle (non-interacting) eigenfunctions in momentum space. Such a single-particle quantum metric has been useful in understanding superconductivity of fermions and Bogoliubov theory of bosons, mostly within mean-field but also with exact approaches~\cite{herzog2022many}, in specific limits such as isolated flat band. It is remarkable that even when a flat band system is strongly interacting by definition (interactions dominate over kinetic energy), mean-field approaches with single particle quantum metric have been relevant, and exact solutions possible -- this boils down to the absence of kinetic energy. It is thus interesting to ask whether the \textit{many-body quantum metric}, which is defined in the parameter space of twisted boundary condition~\cite{Souza:2000PRB,OzawaGoldman2019}, would be important for key physical quantities in flat bands and beyond.

In this work, we investigate the role of the many-body quantum metric on the superfluid properties of Bose systems. We consider one-dimensional interacting lattice models and discover a relation between the superfluid weight, i.e.~the Drude weight, and the many-body quantum metric. We find that the many-body quantum metric enters in the upper bound of the Drude weight for strongly correlated bosonic systems. As a concrete example, we consider one-dimensional Bose-Hubbard Creutz ladder~\cite{creutz}, which hosts two flat bands, and numerically verify the upper bound of the Drude weight for half and unit densities. Our results reveal a connection between the many-body quantum geometry of strongly correlated ground states and their superfluid behaviour. Our predictions can be experimentally tested with current ultracold gas and circuit-QED setups.

\textit{The Drude weight and the many-body quantum metric.---} 
We start by considering a generic one-dimensional lattice Hamiltonian, defined on a ring of length $L$ with periodic boundary conditions
\begin{equation}
    \hat{H}_0 = \hat{H}_\text{kin} + \hat{H}_V + \hat{H}_U = (\hat{K}+\hat{K}^\dagger) + \hat{H}_V + \hat{H}_U,
    \label{OriginalHamiltonian_generic}
\end{equation}
where the kinetic part $\hat{H}_\text{kin}$ is decomposed into hopping along the ring in one direction $\hat{K}$ and that in the opposite direction $\hat{K}^\dagger$. The terms $\hat{H}_V$ and $\hat{H}_U$ are onsite potential term and the inter-particle interaction term, respectively.
By considering an external flux $\Phi$ that threads the ring we implement the twisted boundary conditions. The kinetic part $\hat{H}_\text{kin}$ acquires a phase, while the onsite part $\hat{H}_V$ and the interacting part $\hat{H}_U$ stay invariant.
The Hamiltonian with non-zero flux is then $\hat{H}(\Phi) = \hat{K} e^{i \Phi/L}+\hat{K}^\dagger e^{-i \Phi/L} + \hat{H}_V + \hat{H}_U$, where $\theta = \Phi/L $ is the phase gradient across the 1D ring and is distributed over all the $L$ links.

The superfluid response of the system to such external flux is given by the Drude weight~\cite{kohn1964theory, fye1991drude}
\begin{equation}
    D_w = \pi L \left. \frac{\partial^2 E(\Phi)}{\partial \Phi^2}\right\vert_{\Phi = 0},
    \label{Drude_weight}
\end{equation}
where $E(\Phi)$ represents the ground state energy of $\hat{H}_\Phi$.
The discretized version of Eq.~\eqref{Drude_weight} for a small flux is:
\begin{equation}
        D_w = 2\pi L \frac{E(\Phi) - E(0)}{\Phi^2},
            \label{dw}
\end{equation}
and the energy difference $E(\Phi) - E(0)$ can be evaluated within perturbation theory.
The Hamiltonian~\eqref{OriginalHamiltonian_generic} expanded up to the second order in $\Phi$ is $\hat{H}(\Phi) = \hat{H}_0 + \hat{H}_\text{pert}$, where the small perturbation is
\begin{equation}
\hat{H}_\text{pert} = \frac{\Phi}{L} \hat{J} -\frac{1}{2}\left(\frac{\Phi}{L}\right)^2 \hat{H}_\text{kin},
\label{H_pert}
\end{equation}
having introduced the current operator $\hat{J}= i(\hat{K}-\!\hat{K}^\dagger)$.

If the many-body ground state $ |\Psi_0\rangle $ is non-degenerate at $\Phi=0$, $\hat{H}_0|\Psi_0\rangle = E_0(0) |\Psi_0\rangle$. Applying the second order perturbation theory together with Eq.~\eqref{dw} for the Drude weight, we get \cite{roth2003phase, hetenyi2019superfluid}
\begin{eqnarray}
        D_w &=& \frac{2\pi L}{\Phi^2} \Bigg[
        \frac{\Phi}{L} \langle \Psi_0 | \hat{J}| \Psi_0 \rangle 
        - \frac{\Phi^2}{2L^2} \langle \Psi_0 | \hat{H}_\text{kin} | \Psi_0 \rangle \nonumber
        \\ &&  \qquad \quad \, -\frac{\Phi^2}{L^2}  \sum_{m\neq0} \frac{\langle \Psi_m |\hat{J}| \Psi_0\rangle\langle \Psi_0 |\hat{J}| \Psi_m\rangle}{E_n(0) - E_0(0)} \Bigg]  \nonumber \\ 
        &=&
        - \frac{\pi}{L} \langle \Psi_0 | \hat{H}_\text{kin} | \Psi_0 \rangle -\frac{2 \pi}{L}  \sum_{m\neq0} \frac{|\langle \Psi_m |\hat{J}| \Psi_0\rangle|^2}{E_m(0) - E_0(0)},
        \label{dw_perturbed}
\end{eqnarray}
where we used that $\langle \Psi_0 | \hat{J}| \Psi_0 \rangle $ is zero in the thermodynamic limit as there is no ground state persistent current, and we have only kept terms that survive as $\Phi$ goes to zero.
Equation~\eqref{dw_perturbed} is statistics independent. However, the functional form of the Drude weight will depend on the statistics via the many-body states $|\Psi_0\rangle$. 

The second term in Eq.~\eqref{dw_perturbed}, containing the current operator, can be related to the many-body quantum metric of the Hamiltonian in the twist-angle space. Such many-body quantum metric is defined from the many-body ground state as~\cite{Souza:2000PRB, OzawaGoldman2019}
\begin{equation}
    \mathfrak{g}(\phi) =\mathrm{Re}\left[\langle \partial_\phi \Psi_0 | \left(1- |\Psi_0\rangle \langle \Psi_0| \right) | \partial_\phi \Psi_0\rangle\right]
    \label{metric}
\end{equation}
where $\phi=\Phi/L$.
The many-body quantum metric at $\phi = 0$ can be expressed as
\begin{equation}
    \mathfrak{g}(0) = \mathrm{Re}\left[\sum_{m\neq 0} \frac{|\langle \Psi_m | \hat{J} | \Psi_0 \rangle |^2}{(E_m(0) - E_0(0))^2} \right],
    \label{metric_perturbed}
\end{equation}
having used the following expansion 
$$|\partial_\phi \Psi_0\rangle = \sum_{m\neq 0 } \frac{\langle \Psi_m|\partial_\phi \hat{H}(\phi) | \Psi_0\rangle}{E_m-E_0} |\Psi_n\rangle,$$
together with $\partial_\phi \hat{H}(\phi) = L \partial_\Phi \hat{H}(\Phi)
= \hat{J} + \mathcal{O}(\Phi)$.
Equation~\eqref{metric_perturbed} does not require the knowledge of the eigenstates of the Hamiltonian in the twist-angle space and can be easily connected with the second term appearing in the Drude weight formula in Eq.~\eqref{dw_perturbed}, apart from denominator being squared. In our numerical calculations, we use an alternative, manifestly gauge-invariant, expression of the many-body quantum metric: by writing $|\partial_\phi \Psi_0\rangle \!=\![|\Psi_0(\phi)\rangle - |\Psi_0(0)\rangle]/\phi$ for small enough $\phi$, it can be expressed as 
\begin{equation}
    \mathfrak{g}(0) = \lim_{\phi \rightarrow 0} (1- |\langle \Psi_0(\phi)|\Psi_0(0)\rangle|^2)/\phi^2.
    \label{metric_numeric}    
\end{equation}
Since the many-body quantum metric is defined by a flux inserted via a global twisted boundary condition, it is not dependent on the orbital positions like the single-particle quantum metric. Thus subtleties related to relation between the single-particle quantum metric and superfluid weight~\cite{huhtinen2022revisiting} do not play a role here. 

We now assume that the system has a many-body energy gap $\varepsilon$ so that $E_m - E_0 > \varepsilon$ for any $m \neq 0$. We can then obtain a bound for the second term Eq.\eqref{dw_perturbed} as~\cite{SuppMat}
\begin{equation}
\frac{2\pi}{L} \sum_{m\neq0} \frac{|\langle \Psi_m |\hat{J}| \Psi_0\rangle|^2}{E_m(0) - E_0(0)} > \frac{2\pi}{L} \mathfrak{g}(0) \varepsilon,
\label{bound2}
\end{equation}
which leads to the following upper bound for the Drude weight
\begin{equation}
        D_w < - \frac{\pi}{L}\langle \Psi_0 | \hat{H}_\text{kin} | \Psi_0 \rangle - \frac{2 \pi}{L} \mathfrak{g}\, \varepsilon.
        \label{dw_perturbed_bound}
\end{equation}
This formula is the central result of our work.
We note that generally $\langle \Psi_0 | \hat{H}_\text{kin} | \Psi_0 \rangle < 0$ and thus the first term is positive. The many-body quantum metric $\mathfrak{g}$, defined as in Eq.~\eqref{metric} for $\phi \rightarrow 0$, is a non-negative quantity, and thus the many-body quantum metric tends to \textit{reduce} the upper bound of the Drude weight. This is in contrast to the role of the \textit{single particle} quantum metric defined in momentum space appearing in mean-field descriptions of interacting fermions~\cite{peotta2015superfluidity} and bosons~\cite{julku2021quantum, julku2021excitations}, where the quantum metric tends to enhance the superfluid density.

\begin{figure}[t!]
    \centering
    \includegraphics[width=0.45\textwidth]{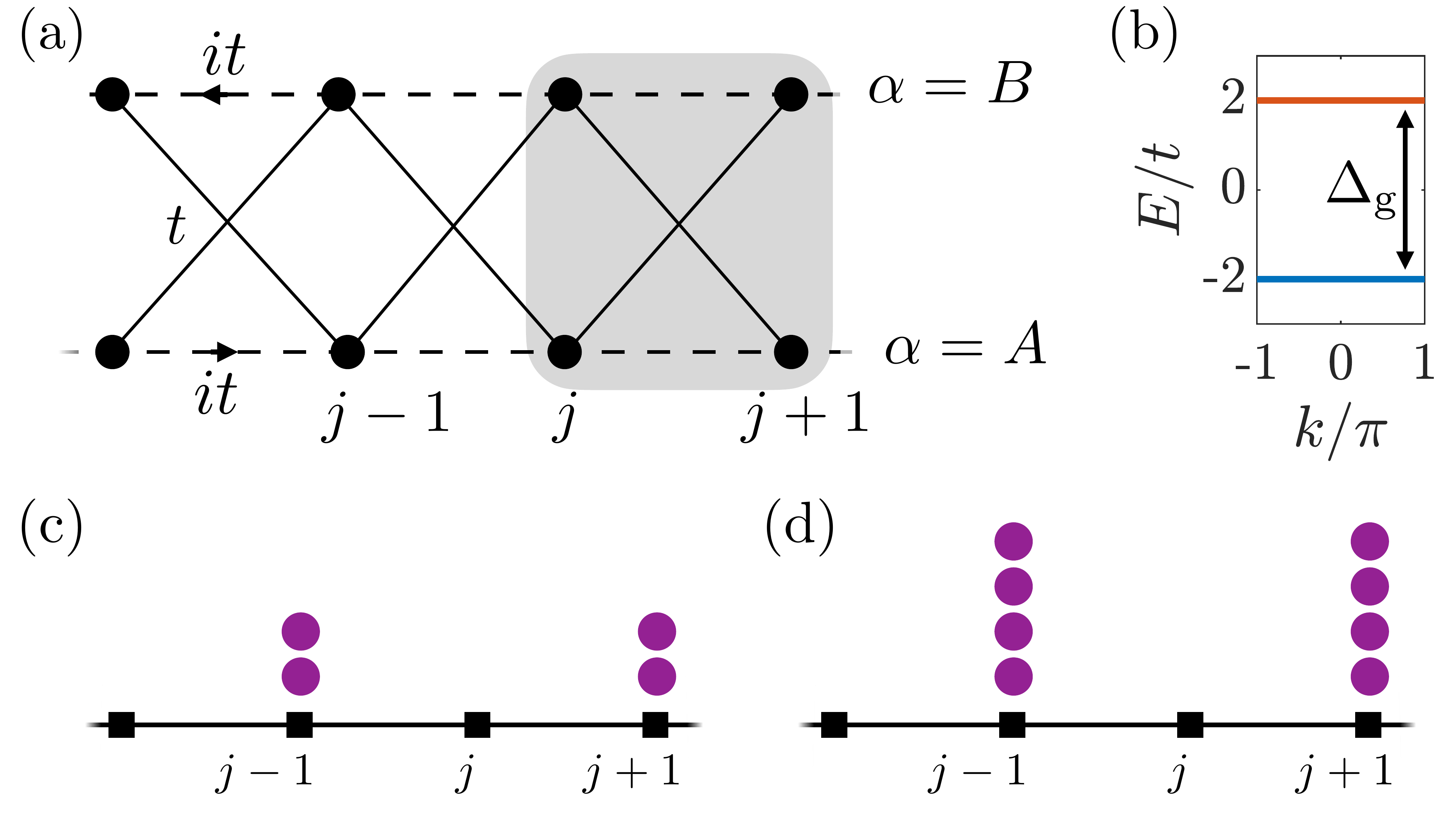}
    \caption{(a) The Creutz ladder with $L=4$ unit cells. Dashed links indicate complex hopping $i t$ following the direction of the arrow; solid links indicate hopping $t$. The flat band state of Eq.~\eqref{flatbandstate} is defined on the plaquette highlighted in grey. (b) The energy dispersion of the non-interacting Creutz ladder, with the two flat bands separated by an energy gap $\Delta_g$. (c-d) The model projected on the lowest flat band in Eq.~\eqref{projectedHam} is defined on $L$ sites, indicated by the black squares. The ground state has a charge-density-wave order for (c) half density and (d) unit density, which are schematically drawn with purple circles denoting bosons. }
    \label{fig:Fig1}
\end{figure}

\textit{The Creutz ladder.---} We now apply our result to the Creutz ladder model, a one-dimensional chain with two orbitals per unit cell on which bosons can hop and interact, see Fig.~\ref{fig:Fig1}(a). The orbitals $\alpha = A, B$ are cross-linked in such a way that a $\pi$-flux is acquired on the plaquette \cite{creutz}. The Hamiltonian with onsite repulsive interactions is~\cite{takayoshi2013phase, tovmasyan2013geometry, tovmasyan2018preformed}
\begin{align}
    \hat{H} =& \sum_{j=1}^L \left(it \hat{a}_{j}^\dagger \hat{a}_{j-1}^{} -it \hat{b}_{j}^\dagger \hat{b}_{j-1}^{} + t \hat{a}_{j}^\dagger \hat{b}_{j-1}^{}+ t \hat{b}_{j}^\dagger \hat{a}_{j-1}^{} + \text{H.c.}\right) \nonumber \\ &+ U \sum_{j=1}^L  \hat{a}_{j}^\dagger \hat{a}_{j}^\dagger \hat{a}_{j}^{} \hat{a}_{j}^{} + U  \sum_{j=1}^L  \hat{b}_{j}^\dagger \hat{b}_{j}^\dagger \hat{b}_{j}^{} \hat{b}_{j}^{}
    \label{OriginalHamiltonian}
\end{align}

The operator $\hat{a}_{j}^\dagger$($\hat{b}_{j}^\dagger$) creates a particle on the $A$($B$) orbital at site $j$, and the periodic boundary condition is implemented by identifying $\hat{a}_0 = \hat{a}_L$ and $\hat{b}_0 = \hat{b}_L$.
In the non-interacting limit ($U=0$), the Creutz ladder has two flat bands at $E=\pm 2t$, separated by a band gap $\Delta_\text{g} = 4 t$, see Fig.~\ref{fig:Fig1}(b). 

The twisted periodic boundary conditions are applied to Eq.~\eqref{OriginalHamiltonian} such that the forward hopping terms acquire a phase $e^{i \Phi/L}$. In the following, the unit cell size is fixed $a = 1$. 
The lowest flat band state has the form
\begin{equation}
    \hat{W}_j^\dagger |0\rangle = \frac 1 2 \left( \hat{a}_j^\dagger + i \hat{b}_j^\dagger - i e^{i \Phi/L} \hat{a}_{j+1}^\dagger - e^{i \Phi/L}\hat{b}_{j+1}^\dagger \right) |0\rangle.
    \label{flatbandstate}
\end{equation}
Such state is obtained from the Wannier functions of the lowest flat band within the twisted periodic boundary conditions~\cite{kohn1959analytic}.
The operator $\hat{W}_j^\dagger$ defines the creation of a boson in the lowest flat band in a state localized around the $j$-th and $j+1$-th unit cells. By neglecting the upper band contributions, we can express the original operators $\hat{a}_i^\dagger$ and $\hat{b}_i^\dagger$ in terms of the lowest flat band operators using the following transformation~\cite{huber2010bose}
\begin{equation}
    \begin{split}
    \hat{a}_i^\dagger =& \frac 1 2  \left(\hat{W}_i^\dagger + i e^{-i \Phi/L} \hat{W}_{i-1}^\dagger\right)\\
    \hat{b}_i^\dagger =& \frac 1 2 \left(-i\hat{W}_i^\dagger - e^{-i \Phi/L}\hat{W}_{i-1}^\dagger\right)
    \end{split}
    \label{transformation_tbc}
\end{equation}
which we use to project the original Hamiltonian in Eq.~\eqref{OriginalHamiltonian} on the flat band states~\cite{tovmasyan2013geometry}
\begin{align}
    \hat{H}(\Phi)&^\text{proj} = \sum_{j=1}^L \Big[\frac U 4 \hat{W}_j^\dagger \hat{W}_j^\dagger \hat{W}_j^{} \hat{W}_j^{} +  \frac U 2 \hat{W}_{j}^\dagger \hat{W}_{j-1}^\dagger \hat{W}_j^{}\hat{W}_{j-1}^{} \nonumber \\& -\frac{U}{8} \left( e^{-2i \Phi/L}\hat{W}_j^\dagger \hat{W}_j^\dagger \hat{W}_{j-1}^{} \hat{W}_{j-1}^{} + \text{H.c.}\right)\Big].
    \label{projectedHam}
\end{align}
This Hamiltonian defines a one-dimensional effective lattice model with $L$ sites (Fig.~\ref{fig:Fig1}(c-d)). The twisted boundary conditions are reflected into the pair hopping term with a phase $\theta=2\Phi/L$, equal to twice the one carried by the single particle. In fact, only pairs of particles can move, while the single particles are localized and interact via the on-site and nearest-neighbour effective interactions. The projected Hamiltonian in Eq.~\eqref{projectedHam} is valid when $U \ll \Delta_\text{g}$ and there is a negligible occupation of the upper flat band. Notice that all terms in Eq.~\eqref{projectedHam} are proportional to $U$, which defines the relevant energy scale of the system. In the following, we will refer to the filling density of particles $n=N_b/N_s$, where $N_b$ is the number of bosons and $N_s=2L$ is the number of sites of the full model.

For half density $n=1/2$, previous works found that the ground state has a charge-density-wave (CDW) order by examining the projected model in Eq.~\eqref{projectedHam} for $\Phi=0$~\cite{takayoshi2013phase, tovmasyan2013geometry}.
We performed the exact diagonalization of the projected model, and confirm that the ground state is the CDW for any value of $\Phi$ as we show in Fig.~\ref{fig:Fig1}(c) and Supplemental Material~\cite{SuppMat}. 
We also find that the energy $E(\Phi)$ is $\pi$-periodic (instead of $2\pi$) with the flux $E(\Phi) = E(\Phi+\pi)$, an indication that the motion is carried by pairs of bosons \cite{tovmasyan2018preformed}. When $L$ is even, we also find that the ground state is two-fold degenerate, where the degeneracy is lifted by finite-size effects, in agreement with Ref.~\cite{takayoshi2013phase}. When $L$ is odd, the ground state for finite $L$ is instead L-fold degenerate, due to a single unpaired particle that can be distributed in the empty sites of the CDW. \footnote{We note that we find a non-negligible discrepancy of the ground state energy as well as the the many-body excitation gap compared to those reported in Ref.~\cite{takayoshi2013phase}, where the projected model is further mapped onto an effective spin-1/2 XXZ model by truncating the Hilbert space to ignore states in which more than two bosons occupy the same site. Our exact diagonalization result shows that the truncation of the Hilbert space cannot be justified; further analysis is given in the Supplemental Material~\cite{SuppMat}.} 

In the $n=1$ unit density case at any flux $\Phi$, the ground state is again two-fold degenerate due to a CDW order of two pairs of bosons, see Fig.~\ref{fig:Fig1}(d). As in the half density case, for odd $L$ the CDW order cannot be hosted in a commensurate way, see Supplemental Material~\cite{SuppMat}. For this reason in the following we will consider only $L$ even both for half and unit density.

To estimate the Drude weight of the Creutz ladder, we expand the projected Hamiltonian in the low-flux limit, paying attention that the flux is $\theta = 2 \Phi/L$ in the pair hopping term upon using Eqns.~\eqref{H_pert}-\eqref{dw_perturbed}-\eqref{dw_perturbed_bound}. The first term of the Drude weight in Eq.~\eqref{dw_perturbed_bound} denotes the kinetic energy of pairs of bosons in the projected model. According to Eq.~\eqref{projectedHam}, when $\Phi = 0$, a pair of bosons hop with an amplitude of $-U/8$ times the bosonic enhancement factor of 2. Its dispersion relation is then $(-U/2) \cos (k)$ where $k$ is the quasimomentum of a pair. The kinetic energy is thus bounded from below by
\begin{equation}
\langle \Psi_0 | \hat{H}_\text{kin} | \Psi_0 \rangle \geq -\frac{U}{2} \cdot \frac{N_b}{2},
\label{bound1_c}    
\end{equation}
where $N_b/2$ is the number of pairs of bosons.

\begin{figure}
    \centering
    \includegraphics[width=0.45\textwidth]{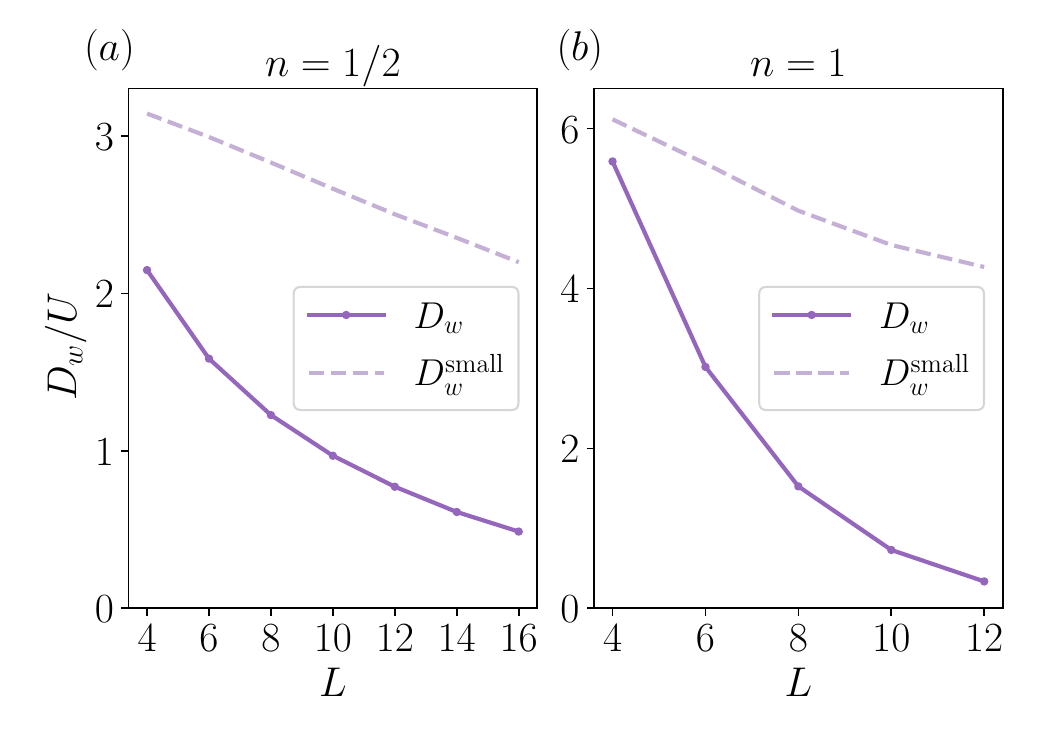}
    \caption{Drude weight of the projected Creutz ladder in units of $U$ as a function of $L$ for filling density $n=1/2$ in (a) and $n=1$ in (b). The solid line is the Drude weight calculated using Eq.~\eqref{dw} within exact diagonalization of the projected Hamiltonian given in Eq.~\eqref{projectedHam}; the upper bound calculated from Eq.~\eqref{dw_perturbed_bound_low} is indicated by the dashed line.}
    \label{fig:Fig2}
\end{figure}

To evaluate the second term of the Drude weight bound in Eq.~\eqref{dw_perturbed_bound}, we find that the first excitations are at $\varepsilon=U/4$ (see Supplemental Material~\cite{SuppMat}). Notice that upon using Eqns.~\eqref{H_pert}-\eqref{dw_perturbed}-\eqref{dw_perturbed_bound} with the flux $\theta = 2 \Phi/L$ carried by the pairs, the $D_w^\text{small}$ gets an extra factor of $4$ with respect to the result with the flux $\theta = \Phi/L$ carried by the single particle.
We then obtain the following upper bound for the Drude weight:
\begin{equation}
        D_w^\text{small} <  \pi U \left(2 n - \frac{2\mathfrak{g}^{\text{proj}}(0)}{L}\right),
        \label{dw_perturbed_bound_low}
\end{equation}
where we introduced the many-body quantum metric $\mathfrak{g}^{\text{proj}}$ of the projected Hamiltonian. We note that $\mathfrak{g}^{\text{proj}}$ can be in general different from the many-body quantum metric of the full Hamiltonian $\mathfrak{g}$, as it is constructed from the states in a reduced (projected) space in the isolated flat band limit. 

The Drude weight and the bound in~\eqref{dw_perturbed_bound_low} are linear with the interaction $U$, as long as $U \ll \Delta_\text{g}$ and the projected Hamiltonian approach is valid. The Drude weight, as calculated from Eq.~\eqref{dw}, is plotted in units of $U$ as a function of $L$ for filling density $n=1/2$ in Fig.~\ref{fig:Fig2}(a) and for filling density $n=1$ in Fig.~\ref{fig:Fig2}(b). We thus numerically confirm that the formula Eq.~\eqref{dw_perturbed_bound_low} indeed provides an upper bound.

We now briefly discuss the large interaction limit, for which there are non-negligible upper-band contributions to the flat-band projection Eq.~\eqref{transformation_tbc}. In the limit $U \geq \Delta_\text{g}$, we need to use the full Hamiltonian in Eq.~\eqref{OriginalHamiltonian} with twisted periodic boundary conditions to derive the Drude weight in Eq.~\eqref{dw_perturbed}. 

\begin{figure}
    \centering
    \includegraphics[width=0.45\textwidth]{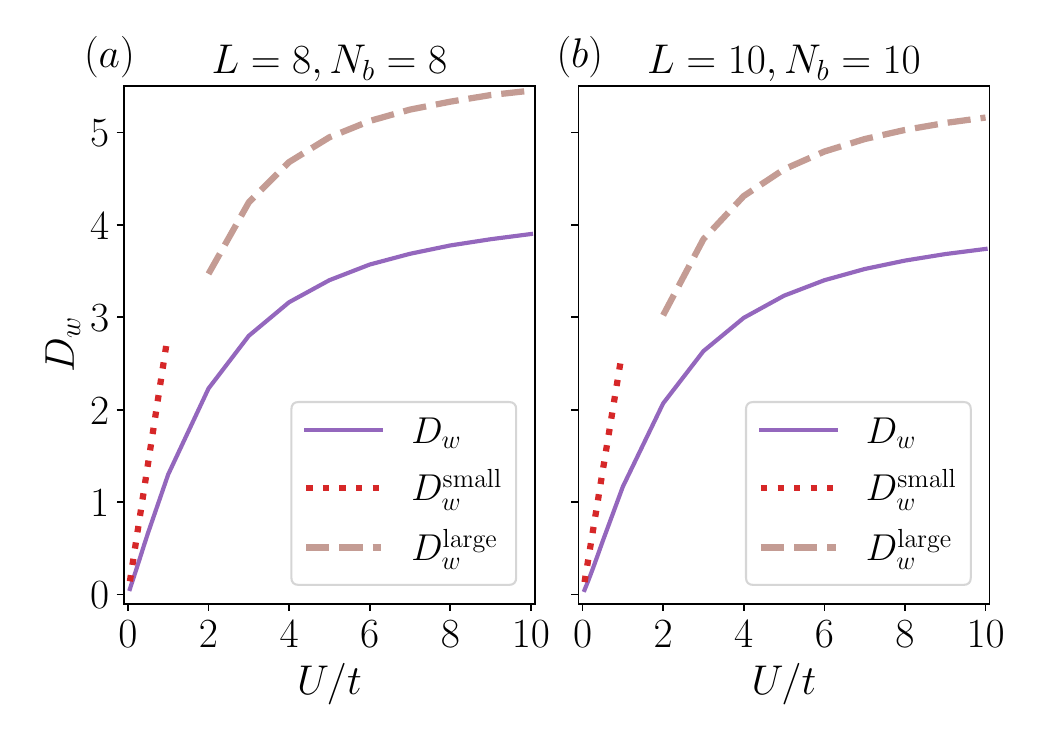}
    \caption{Drude weight of the full (non-projected) Creutz ladder as a function of $U$ for half filling $n=1/2$. The solid line is the Drude weight calculated using Eq.~\eqref{dw} from the exact diagonalization of the full Hamiltonian Eq.~\eqref{OriginalHamiltonian} with twisted boundary conditions for $L=8$ (a) and $L=10$ (b). The bound calculated from Eq.~\eqref{dw_perturbed_bound_large} for $U>2t$ is indicated by the dashed line. For comparison, the bound calculated from Eq.~\eqref{dw_perturbed_bound_low} is also indicated with a dotted line for $U<t$.}
    \label{fig:Fig3}
\end{figure}

For half density, the relevant energy gap is the band gap $\varepsilon = \Delta_g = 4t$, since now the excitations of the ground state in a completely filled lower band involve states where at least one particle is in the higher band, rather than two pairs occupying the same site. The kinetic energy in the ground state is bounded by
$$\langle \Psi_0 | \hat{H}_\text{kin} | \Psi_0 \rangle \geq - 2 t N_b.$$
The bound in Eq.~\eqref{dw_perturbed_bound} then becomes 
\begin{equation}
        D_w^{\text{large}} <  4\pi t \left(n- \frac{2 \mathfrak{g}(0)}{L}\right).
        \label{dw_perturbed_bound_large}
\end{equation}
The many-body quantum metric $\mathfrak{g}$ is now defined for the full Hamiltonian in Eq.~\eqref{OriginalHamiltonian} with twisted boundary conditions using Eq.~\eqref{metric}. In Fig.~\ref{fig:Fig3} the Drude weight for the half density is shown as a function of the interaction strength; we numerically confirm both the upper bound for the Drude weight given by the inequality~\eqref{dw_perturbed_bound_large} for large interactions as well as the one for small interactions, Eq.~\eqref{dw_perturbed_bound_low} .

For unit density at large interactions, the Mott insulator phase becomes the ground state. The Drude weight then approaches zero at large interactions and the upper bound (\ref{dw_perturbed_bound_large}) tends to overestimate the Drude weight (see Supplemental Material~\cite{SuppMat} for more details).

\textit{Discussion and conclusions.---} 
We have studied relations between the Drude weight and the many-body quantum metric of one-dimensional bosonic systems. The role of the \textit{many-body} quantum metric, which is defined in the parameter space of the twist-angles, turned out to be opposite to the role of the \textit{single-particle} quantum metric defined in momentum space: here we find the former to \textit{reduce an upper bound} while previous works related the latter to \textit{enhancement} of the superfluid density. Large many-body quantum metric means that the eigenstates of the interacting system deviate strongly when a flux is inserted, thus intuitively leading to small response, i.e.~smaller superfluid weight. In contrast, in the mean-field and exact treatments of superfluid fermions~\cite{peotta2015superfluidity,torma2018quantum,huhtinen2022revisiting,herzog2022many}, the single particle quantum metric does not describe the actual state of the interacting system, however, its large value means strong delocalization and overlap of the single particle Wannier functions that allow effective pair hopping in the system. It would be intriguing to study the role of the many-body quantum metric in fermionic systems in more detail. However, since the bound (\ref{dw_perturbed_bound}) is independent of particle statistics, we can already at this point make the intriguing conclusion that superconductivity of fermionic (Cooper) pairs is upper bounded by the many-body quantum metric.

The bound (\ref{dw_perturbed_bound}) is independent of spatial dimensions too. It is of great interest to extend the analysis to two or higher dimensions. In two dimensions, an inequality connecting the many-body quantum metric and the many-body Chern number is known. With that inequality, one may be able to find the upper bound of the Drude weight in terms of the many-body Chern number, which is relevant, for example, in studying fractional quantum Hall effects. 

Our work shows that many-body quantum geometry can play an important role in understanding transport properties of interacting systems, inspiring further theoretical and experimental studies. The Creutz ladder has been experimentally realized in ultracold atomic gases~\cite{Junemann:PRX2017,Kan:2020NJP} through resonantly modulated optical lattices. A similar technique is also applicable in circuit QED and other quantum engineered systems to realize all-bands-flat physics~\cite{alaeian2019creating, hung2021quantum, martinez2023interaction}. There are also proposals to measure the many-body quantum metric by applying oscillating forces and detecting the excitations of the system~\cite{OzawaGoldman2018,OzawaGoldman2019}. 
The fundamental relation between the Drude weight and the many-body quantum metric that we predicted can be tested in such experimental platforms.

\begin{acknowledgments}
We thank Ville Pyykk\"{o}nen for useful discussions. The calculations have been performed using the QuSpin package~\cite{Quspin1, Quspin2}. We acknowledge the computational resources provided by the Aalto Science-IT project. This work was supported by Academy of Finland under Projects No. 303351, No. 327293, and
No. 349313. GS has received funding from the European Union's Horizon 2020 research and innovation programme under the Marie Sk\l{}odowska-Curie grant agreement No 101025211 (TEBLA).
TO is supported by JSPS KAKENHI Grant No. JP20H01845 and JST CREST Grant No.JPMJCR19T1. This work has been initiated and supported by the Global Intellectual Incubation and Integration Laboratory (GI$^3$ Lab) Program provided by WPI-AIMR, Tohoku University.
\end{acknowledgments}



\section*{Supplementary material}

# Drude weight bounded by the many-body quantum metric in one-dimensional bosonic flat band systems

G. Salerno,[1,*] T. Ozawa,[2] and P. Törmä[1,2,†]

[1]*Department of Applied Physics, Aalto University School of Science, P.O. Box 15100, Aalto, FI-00076, Finland*
[2]*Advanced Institute for Materials Research (WPI-AIMR), Tohoku University, Sendai 980-8577, Japan*

## Supplemental Material

### TRUNCATION OF THE HILBERT SPACE

For small systems, the exact diagonalization is done by considering all states in the Hilbert space of the projected Hamiltonian in Eq. (14) of the main text. However, due to the exponential increase of the number of states, suitable truncation has to be performed to allow reasonable computational effort for larger sized systems. The following results were utilized for estimating what is an appropriate truncation of the Hilbert space for the exact diagonalization on

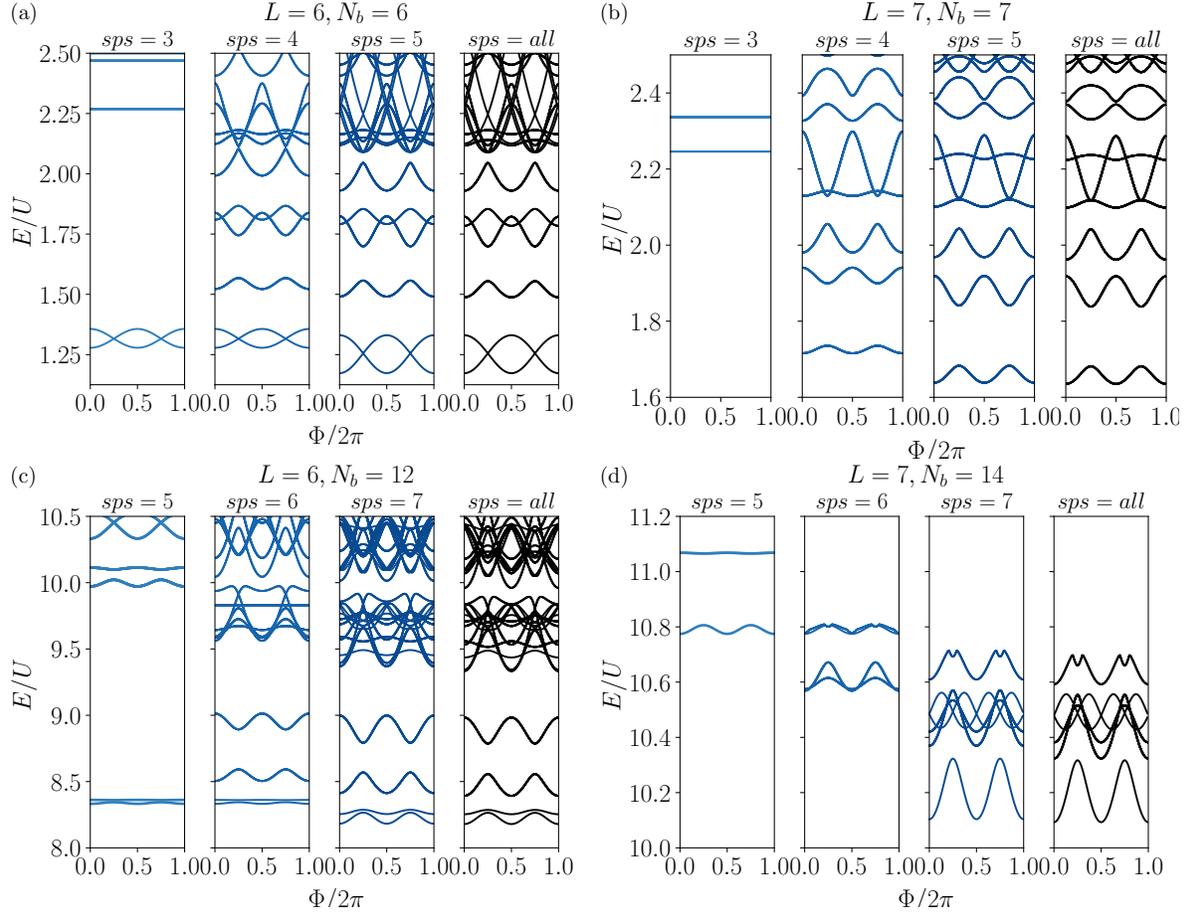

FIG. S1. Low-energy states of the Hamiltonian in Eq. (14) of the main text as a function of the inserted flux $\Phi$, where the Hilbert space has been truncated to a given state-per-site (sps) (maximum number of bosons on a site plus one (for zero occupation). (a) Half density case for $L = N_b = 6$. To obtain the correct ground state energy, the number of bosons in a single site has to be at least 4 (sps=5). (b) Half density for $L = N_b = 7$. The ground state is 7-fold degenerate. (c) Unit density for $L = 6, N_b = 12$. To obtain the correct ground state energy, the number of bosons in a single site has to be at least 6 (sps=7). (d) Unit density for $L = 7, N_b = 14$, showing only the first 20 states. The ground state is singly degenerate.

ladder of lengths larger than $L > 7$ in the main text.

A previous work by Takayoshi et al. [S1] has neglected states in which more than two bosons occupy a single site, namely considering only bosonic occupation of 0, 1 or 2 and thus truncating the Hilbert space to 3 state-per-site (sps). Although this approximation seems justified by the on-site repulsion in the projected Hamiltonian Eq. (14) of the main text, our exact diagonalization calculations show that such truncation is too strict and does not reproduce correctly the ground state energy of the full Hilbert space, see Fig. S1. We have considered both even and odd lengths at half density, respectively $L = N_b = 6$ in Fig. S1(a) and $L = N_b = 7$ in Fig. S1(b). We see that at least occupation of 4 bosons per site (sps= 5) must be considered to recover the correct ground state energy. As mentioned in the main text, the ground state in the thermodynamic limit is a charge-density-wave (CDW) of pair bosons [S1], and the degeneracy is lifted by finite-size effects in Fig. S1(a). Allowing bosonic occupation of 0, 1, 2, 3 or 4 implies that we are retaining states where a pair can hop on the CDW background. For $L$ odd, the ground state is $L$-fold degenerate, due to a single unpaired particle that can be distributed in the empty sites of the CDW in Fig. S1(b).

For unit density $L = 6, N_b = 12$ in Fig. S1(c), and $L = 7, N_b = 14$ in Fig. S1(d) an occupation of at least three pairs on the same site (sps= 7) has to be considered in the truncation to recover the energy of the full Hilbert space. Due to the nature of the ground state, such truncation retains a state where a pair can hop on the CDW background made of four bosons. As in the half density case, for odd $L$ the CDW order is disrupted by an additional pair that can be distributed in the empty sites of the CDW. Most notably, in this case a ground state where all the sites are uniformly occupied by a pair of particles has a larger energy due to the nearest-neighbour interaction in the projected Hamiltonian Eq. (14) of the main text.

## CORRELATION FUNCTIONS

From the exact ground state, we calculated the density-density correlation function

$$g^{(2)}(r) = \frac{1}{L} \sum_r \frac{\langle \hat{W}_0^\dagger \hat{W}_0 \hat{W}_r^\dagger \hat{W}_r \rangle}{\langle \hat{W}_0^\dagger \hat{W}_0 \rangle \langle \hat{W}_r^\dagger \hat{W}_r \rangle} \tag{S1}$$

where $\langle \cdot \rangle$ is the expectation value on the ground state. These correlations functions are plotted in Fig. S2 for half and unit filling, for $L$ even. The $g^{(2)}(r)$ correlations show an oscillating behavior, arising from the spatial modulation of the density in the ground state, where the periodicity of $r = 2$ is related to the CDW pattern.

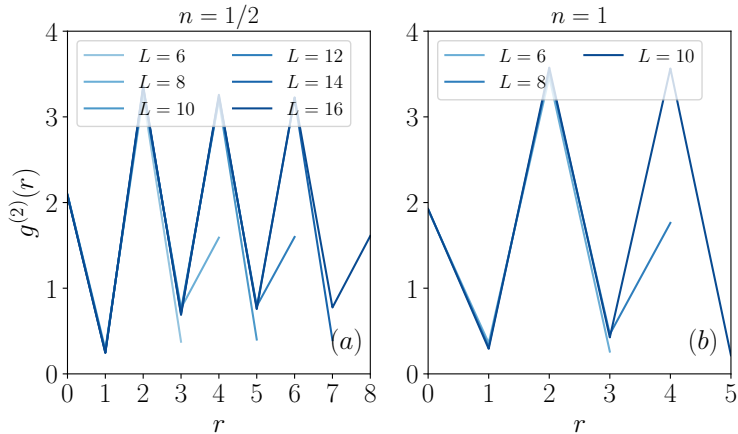

FIG. S2. Density-density correlations in the ground state for (a) half and (b) unit density at various lengths $L$. The oscillations are related to the CDW order.

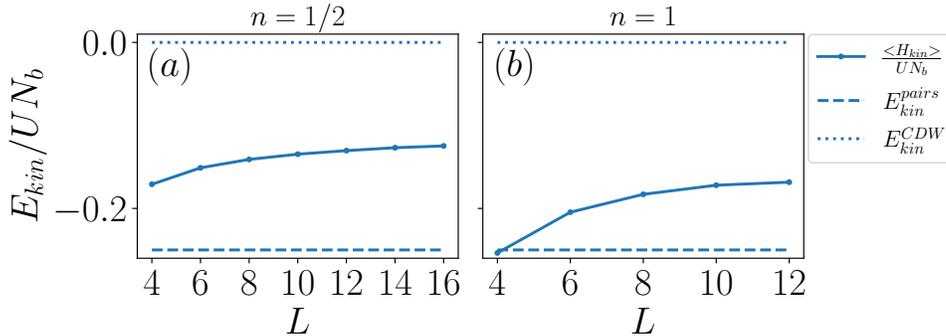

FIG. S3. Kinetic energy of the ground state for (a) half and (b) unit density at various lengths $L$ in units of the interaction strength and the total number of bosons. The lower bound is $E_kin^{pairs} = -UN_b/4$, while the upper bound is $E_{kin}^{CDW} = 0$.

## CREUTZ LADDER GROUND STATE KINETIC ENERGY

We provide a more detailed discussion of the kinetic energy bound given in Eq. (15) of the main text for the projected model of the Creutz ladder.

From Eq. (14) of the main text, the hopping amplitude from moving a single pair from site $j$ to site $i$ is $-\frac{U}{8}\langle 2,0|\hat{W}_i^\dagger \hat{W}_i^\dagger \hat{W}_j \hat{W}_j|0,2\rangle = -\frac{U}{4}$, while the energy dispersion is $(-\frac{U}{4})2\cos(k)$, where $k$ is the pair quasimomentum. In the ground state for $k=0$, a single pair has kinetic energy $-U/2$, while for $N_b$ bosons $E_{kin}^{pairs} = -UN_b/4$, where $N_b/2$ is the number of pairs. Such value represents the maximum absolute value of the kinetic energy, corresponding to the energy of a dilute pair superfluid. Notice that when the density is increased there are additional processes, e.g. when moving one pair from a doubly-pair-occupied site $-\frac{U}{8}\langle 2,2|\hat{W}_i^\dagger \hat{W}_i^\dagger \hat{W}_j \hat{W}_j|0,4\rangle = -\frac{U}{4}\sqrt{6}$, which contribute to lower even more the kinetic energy of the pair superfluid. From Fig. S3, we see that already $E_{kin}^{pairs} = -UN_b/4$ is a good approximation for both $n=1/2$ and $n=1$, which was used in Eq. (15) of the main text.

If we consider a pure CDW-type ground state, such as the one in Fig.(1)(c-d) of the main text, the kinetic energy is strictly $E_{kin}^{CDW} = 0$, as no single (pair or other) hopping process are able to act on the CDW ground state. From Fig. S3, we see that the kinetic energy of the ground state is $E_{kin}^{pairs} \leq \langle H_{kin}\rangle \leq E_{kin}^{CDW}$, meaning that the system lowers the energy by deviating from a pure CDW to allow particles to become superfluid.

## DISCUSSION ON EQUATION (9) OF THE MAIN TEXT

We now discuss how the energy gap $\varepsilon$ in Eq. (9) is evaluated for the Creutz ladder. In Fig. S4 we show the energy of the projected Hamiltonian in Eq. (14) of the main text in units of $U$ as a function of the inserted flux $\Phi$ for half filling density at $L = 6, 8, 10, 12$. The dashed line indicates a bound for the energies of the first excitations $\varepsilon$ in units of $U$, where $\varepsilon = U/4$ is used to evaluate Eq. (16) in the main text. The two lowest energies correspond to the two CDW states, which in the thermodynamic limit are degenerate.

Figure S5 shows the Drude weight component that is proportional to the quantum metric for the projected model $D_w^{(2)} = D_w^{\text{small}} + \frac{4}{\pi}L\langle\Psi_0|\hat{H}_{\text{kin}}|\Psi_0\rangle$ from Eq. (10) of the main text, paying attention to the pair hopping phase $\theta = 2\Phi/L$. The term $D_w^{(2)}$ is compared to its bound in Eq. (16) of the main text, which has to satisfy the following inequality $D_w^{(2)}/U < \pi\frac{2\mathfrak{g}^{\text{proj}}(0)}{L}$

## THE DRUDE WEIGHT FOR UNIT DENSITY AT LARGE INTERACTIONS

We finally discuss the large interaction limit, where for unit density the ground state is a Mott insulator. Figure S6 shows that the Drude weight approaches zero at large interactions and, although the upper bound in Eq. (17) of the main text is still an upper bound, it tends to overestimate the Drude weight.

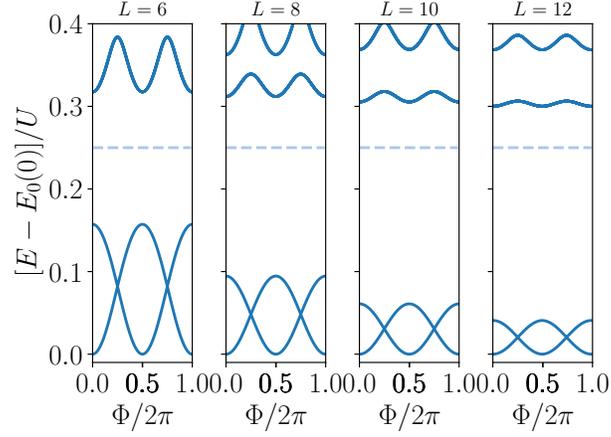

FIG. S4. Low-energy states of the Hamiltonian in Eq. (14) of the main text in units of $U$ as a function of the inserted flux $\Phi$ for half filling density and various $L$. The dashed line indicates a bound for the energies of the first excitations $\varepsilon$ in units of $U$, where $\varepsilon = U/4$.

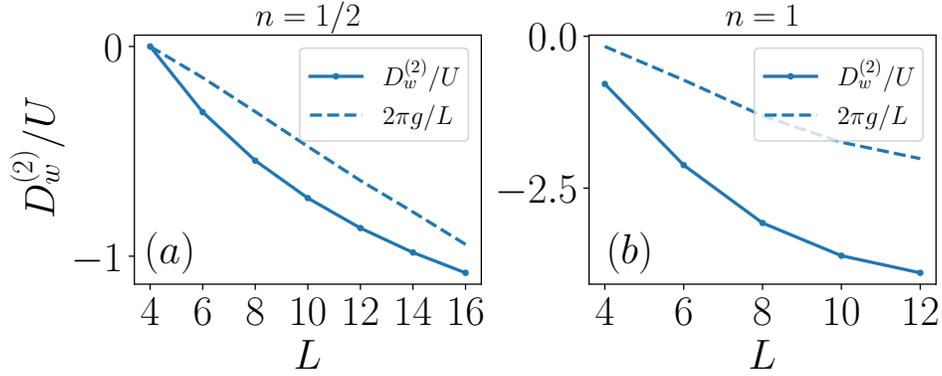

FIG. S5. The Drude weight component proportional to the quantum metric in Eq. (9) of the main text at (a) half and (b) unit density at various lengths $L$.

* grazia.salerno@aalto.fi  
† paivi.torma@aalto.fi

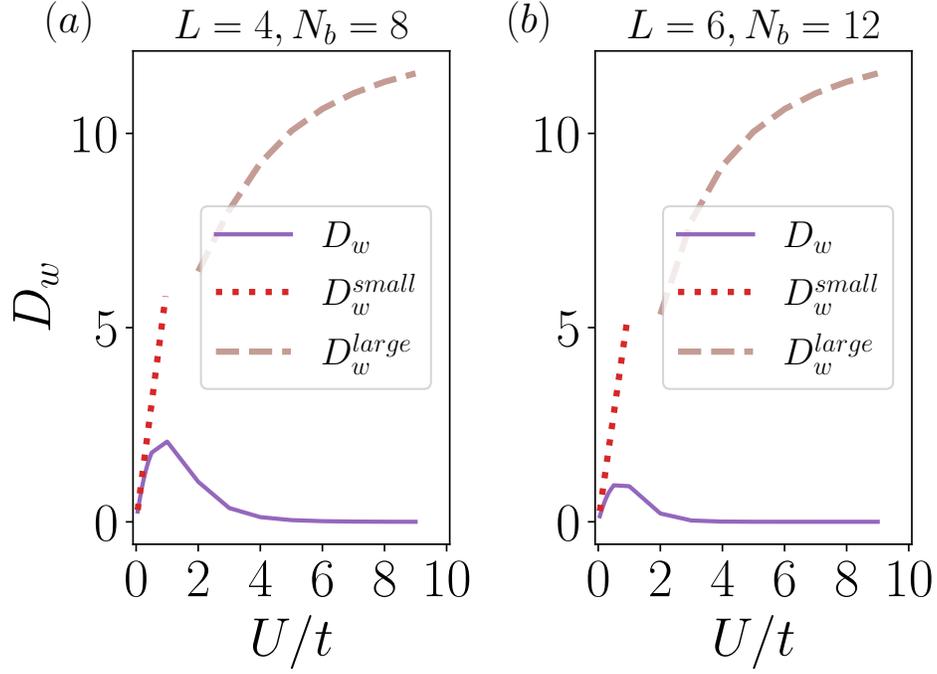

FIG. S6. Drude weight of the full (non-projected) Creutz ladder as a function of $U$ for unit filling $n = 1$. The solid line is the Drude weight calculated using Eq. (3) of the main text from the exact diagonalization of the full Hamiltonian Eq. (11) of the main text with twisted boundary conditions for $L = 4$ (a) and $L = 6$ (b). The bound calculated from Eq. (17) of the main text for $U > 2t$ is indicated by the dashed line. For comparison, the bound calculated from Eq. (16) of the main text is also indicated with a dotted line for $U < t$.


\end{document}